\documentclass[%
 reprint,
 superscriptaddress,
nofootinbib,
 amsmath,amssymb,
 aps,
]{revtex4-2}

\usepackage{graphicx}
\usepackage{dcolumn}
\usepackage{bm}
\usepackage{hhline}
\usepackage{xcolor}
\usepackage{hyperref}
\usepackage{orcidlink}
\usepackage{graphicx}
\usepackage{subcaption}
\usepackage{units}
\usepackage{booktabs}

\newcommand{\imrphenomXPHM}{\textsc{IMRPhenomXPHM}~}
\newcommand{\chieff}{\chi_{\mathrm{eff}}}
\newcommand{\imrphenompv}{\textsc{IMRPhenomPv2}}

\ExplSyntaxOn
\NewDocumentCommand{\longdash}{ O{2} }
 {
  --\prg_replicate:nn { #1 - 1 } { \negthinspace -- }
 }
\ExplSyntaxOff

\newcounter{RunIDCounter}

\definecolor{shgreen}{rgb}{0.15625, 0.609375, 0.316406}

\newcommand{\bw}[0]{\textsc{bayeswave}}
\newcommand{\bilby}[0]{\textsc{bilby}}
\newcommand{\tbilby}[0]{\textsc{tBilby}}
\newcommand{\dynesty}[0]{\textsc{dynesty}}
\newcommand{\bg}[0]{\textsc{bilby}\_\textsc{glitch}}

\newcommand{\SPA}{School of Physics and Astronomy, Monash University, Clayton VIC 3800, Australia}
\newcommand{\OzGravMonash}{OzGrav: The ARC Centre of Excellence for Gravitational Wave Discovery, Clayton VIC 3800, Australia}

\begin{document}


\title{Joint inference for gravitational-wave signal and noise glitch: Method and application}

\author{Shun Yin Cheung~\orcidlink{0009-0009-6918-5764}}
\email{shun.cheung@monash.edu}
\affiliation{\SPA}
\affiliation{\OzGravMonash}
\author{Rhiannon Udall~\orcidlink{0000-0001-6877-3278}}
\affiliation{Department of Physics and Astronomy, University of British Columbia, Vancouver,
British Columbia, V6T1Z4, Canada}

\author{Derek Davis~\orcidlink{0000-0001-5620-6751}}
\affiliation{Department of Physics, University of Rhode Island, Kingston, RI 02881, USA}


\author{Paul D. Lasky~\orcidlink{0000-0003-3763-1386}}
\affiliation{\SPA}
\affiliation{\OzGravMonash}
\author{Eric Thrane~\orcidlink{0000-0002-4418-3895}}
\affiliation{\SPA}
\affiliation{\OzGravMonash}

\date{\today}

\begin{abstract}
\noindent
Non-Gaussian noise transients (``glitches'') in gravitational-wave observatories degrade our ability to accurately perform astrophysical inference.
We present the analysis pipeline \bg, which allows for simultaneous Bayesian inference of gravitational-wave signals and glitches.
Our framework is modular and built on top of the popular \bilby~framework, facilitating future extensions with additional glitch and signal models.
We integrate transdimensional \bilby~into our framework and discuss three glitch models: a physically-motivated slow scattering model, and flexible sine-Gaussian and chirplet models. 
Using a combination of simulated and real data, we demonstrate that \bg~produces reliable results.
We then reanalyse two gravitational-wave events---GW191109 and GW200129---which show signs of interesting black-hole spins, but which may also be affected by data-quality issues.
Our results for GW191109 are consistent with previous analysis.
For GW200129, we recover results consistent with [Payne \textit{et al.}, Phys. Rev. D \textbf{106}, 104017 (2022)
], where the evidence of spin-precession is much weaker when using the waveform approximant \textsc{NRSur7dq4} in combination with wavelet-based glitch modeling.
Furthermore, we show the evidence of spin precession of this event is dependent on the interplay between the waveform approximant and glitch model, since in contrast to \textsc{NRSur7dq4} we find that inference with the waveform approximant \imrphenomXPHM shows strong evidence of spin-precession when used in combination with wavelet-based glitch modeling.

\end{abstract}

\maketitle

\section{Introduction}

In the decade since the detection of the first gravitational-wave (GW) signals, the field has seen myriad advances in our understanding of the mergers of black holes and neutron stars. These include precise characterisation of hundreds of individual events~\cite{GW150914_params, GW170817_paper, LIGOScientific:2025wao, LIGOScientific:2025cmm, LIGOScientific:2025brd, LIGOScientific:2025slb, KAGRA:2021vkt, LIGOScientific:2021usb}, inferences of population properties~\cite{pop_GWTC2, LIGOScientific:2025pvj, KAGRA:2021duu}, 
and tests for physics beyond general relativity \citep{TGR_GWTC1, TGR_GWTC2, TGR_GWTC3, LIGOScientific:2026fcf, LIGOScientific:2026qni, LIGOScientific:2026wpt}. 
All of these studies rely upon the machinery of Bayesian inference~\cite{bilby_paper, Veitch:2014wba, yelikar_low-latency_2023, Dax:2021tsq, Cornish:2014kda}, which in turn rely on assumptions about the behavior of the underlying noise processes.
These assumptions are broken by the presence of non-stationary, non-Gaussian noise transients called ``glitches''~\cite{Davis:2022dnd, LIGO:2024kkz, Virgo:2022ysc} which occur approximately once per minute \cite{GWTC-3_paper}.
Glitches can confound search pipelines~\cite{Usman:2015kfa, Sachdev:2019vvd, LIGO:2024kkz} and bias parameter estimation when they overlap with true GW events~\cite{GW170817_paper, Davis:2022ird, Payne:2022spz, Udall_2025,Udall:2025bts, Gupte:2026whi, Ghonge_2024, Lecoeuche:2026aix, Hourihane:2025vxc}.

The astrophysical interpretation of GW events can be altered by the presence of glitches if they are not properly mitigated.
Famously, the binary neutron star merger GW170817 featured a loud glitch approximately $1.1~\text{s}$ before the merger which complicated coincident detection and accurate sky localisation \cite{GW170817_paper, Pankow:2018qpo}. 
The binary black hole merger GW191109\_010717 (hereafter GW191109) \cite{GWTC-3_paper} may, in the worst-case scenario, have the completely wrong effective aligned spin measurement due to the presence of an unmitigated glitch, although the balance of evidence suggests a less dire outcome~\cite{Udall_2025}.
The evidence of spin-precession in GW200129\_065458 (hereafter GW200129; \citet{GWTC-3_paper}) has also been shown to depend on the glitch model used with some studies suggesting that the black holes exhibit from maximal in-plane spin \cite{Hannam:2021pit} and other studies finding no evidence for precession~\cite{Payne:2022spz}. 
Evidence for other effects such as eccentricity~\cite{Gupte:2026whi} and GW memory~\cite{Cheung_2024} can also be significantly complicated by the effect of low-frequency glitches.

Three broad approaches exist to mitigate the impact of glitches: removing the affected data, subtracting the glitch from the strain data, or incorporating the modeling of the glitch into the inference algorithm itself. 
The first of these approaches, cutting out affected data in either time or frequency domain \citep{Pankow:2018qpo, Davis:2022ird, Payne:2022spz, Udall_2025}, reliably removes the effects of the glitch, but in doing so sacrifices astrophysical information. 
The second approach consists of modeling the glitch with some algorithm~\cite[e.g.,][]{Cornish:2020dwh, Ghonge_2024, bondarescu_2023}, often itself a Bayesian inference algorithm, then subtracting the resulting modeled reconstruction of the glitch from the original strain data, before proceeding with inference as usual. 
An implementation of this method~\cite{Cornish:2020dwh} in the~\bw{}~code \cite{Cornish:2014kda} is the approach currently adopted by LIGO-Virgo-KAGRA (LVK)\cite{ligo_2009, virgo_2012, KAGRA_2013} collaboration analyses~\citep{LIGOScientific:2021usb, KAGRA:2021duu, LIGOScientific:2025slb,  Davis:2022ird}.
However, it has been shown that in a number of cases analyses which use the subtraction method may still be impacted by residual glitch power~\cite{Payne:2022spz, Udall_2025, Gupte:2024jfe, Gupte:2026whi,Udall:2025bts}. 

A more robust method for glitch mitigation is to perform inference with a signal and glitch model simultaneously \citep{Chatziioannou:2021ezd, Hourihane:2022doe, Udall_2025, malz_2025}.
This approach is implemented in \bw{}~\citep{Chatziioannou:2021ezd, Hourihane:2022doe}, which allows for joint signal and glitch inference as well as glitch subtraction using a transdimensional Morlet-Gabor wavelet model. \citet{Udall_2025} demonstrated joint inference with a signal and glitch with a scattered light model, using the precursor to the \bg~ implementation discussed in this work. 
Joint inference has been successfully applied to GW170817~\cite{Chatziioannou:2021ezd},  GW200129~\cite{Payne:2022spz}, GW191109~\cite{Udall_2025}, and approximated for GW190701\_203306, GW231114\_043211, and GW231223\_032836~\cite{Gupte:2026whi}.

Another option which has seen increasing interest is to modify the likelihood function, or forego the use of a likelihood function at all. 
For example, \citet{Ashton:2022ztk} models noise with a Gaussian process to capture the behavior of both the glitch and the Gaussian noise . 
Yet another approach is to use machine learning to model the full noise distribution---including any transient artifacts---and hence allow simultaneous inference of the true underlying signal~\cite{Legin_2025,emma_2026,Chatterjee_2024}.

Despite the robustness of joint signal and glitch inference, such approaches are not routinely incorporated in standard GW parameter estimation pipelines. 
Instead, glitch mitigation is applied on a case-by-case basis, only after a transient noise artifact is identified. This is largely due to the added computational cost and the complexity of its application. Nevertheless, incorporating glitch mitigation directly into standard parameter estimation pipelines will lead to more robust GW inference. 

In this paper, we present \bg \footnote{The code can be found on the \textsc{git} repository \hyperlink{https://git.ligo.org/rhiannon.udall/bilbyparametricglitch}{https://git.ligo.org/rhiannon.udall/bilbyparametricglitch}}, a code package which performs joint inference between a GW signal and a glitch that we make publicly available.
It is an extension to the widely used Bayesian inference library \bilby~\citep{bilby_paper}. By building on the existing \bilby~infrastructure, we enable \bilby~users to perform glitch mitigation within the standard GW inference workflows without the need to adopt a separate analysis framework.
\bg~ is modular, allowing additional glitch models to be incorporated over time to study and mitigate different types of glitches.
The current implementation includes a slow scattered light model, the more flexible Sine-Gaussian wavelet model used by \bw{}~\cite{Cornish:2014kda} and the chirplet model introduced by~\citet{Millhouse:2018dgi}.
The analysis pipeline supports flexible glitch modeling by incorporating transdimensional sampling using \tbilby~\citep{tbilby_paper}, allowing the number of components in the glitch model to be inferred directly from the data.

The method implemented in our code is described in Section~\ref{sec:method}, with the glitch models described in Section~\ref{sec:models}. We validate pipeline performance in \ref{sec:valid} and apply the pipeline to GW events GW191109 and GW200129 in Section \ref{sec:app}. We present concluding remarks and direction for future work in Section \ref{sec:conclusion}.

\section{Methodology}\label{sec:method}
The objective of GW inference is to construct a posterior distribution of the parameters of the binary $\theta$. For given data $d$, the posterior can be calculated using Bayes theorem, 
\begin{equation}
    p(\theta|d) = \frac{\mathcal{L}(d|\theta)\pi(\theta)}{\mathcal{Z}(d)},
\end{equation}
where $\mathcal{L}(d|\theta)$ is the likelihood, $\pi(\theta)$ is the prior and $\mathcal{Z}(d)$ is the Bayesian evidence.

The standard Whittle likelihood used in GW inference~\cite{Veitch:2014wba} assumes noise to be stationary and Gaussian
\begin{equation}
    \ln \mathcal{L}(d|\theta) = -\frac{1}{2}\langle d-h(\theta), d-h(\theta)\rangle + \text{const} ,
\end{equation}
where $h$ is the signal waveform, and we define the noise-weighted inner product as 
\begin{equation}\langle x, y \rangle \equiv 4\Delta f\sum_i\text{Re}\left[\frac{\tilde{x}^*_i\tilde{y}_i}{P_i}\right],
\end{equation}
where $\Delta f$ is the frequency resolution, $\tilde{x}$ represents the Fourier-domain representation of $x$, and $P$ is the noise power-spectral density \cite{Thrane:2018qnx}.


In order to model the glitch and signal waveform simultaneously, we assume that the data $d(t)$ consists of the linear sum of these two models along with stationary Gaussian noise $n(t)$, such that
\begin{equation}
    d = n + g(\gamma) + h( \theta),
\end{equation}
where $g(\gamma)$ and $h(\theta)$ represent the glitch and signal waveform parameterised by $\gamma$ and $\theta$, respectively.
The standard Whittle likelihood is then modified to replace $h(\theta)$ with $h(\theta)+g(\gamma)$.

The inference can be computationally expensive given the combined dimensionality of both signal and glitch parameters. Therefore, we reduce the dimensionality of the problem by analytically marginalising over the the luminosity distance, extending the method already implemented in \bilby~to the joint likelihood case (see Appendix~\ref{sec:distance-marg}).

In the next Section we introduce a number of different glitch models $g(\gamma)$ that are implemented in \bg. 
For each of these models, the number of glitch parameters $\gamma$ is not known \textit{a priori}, suggesting the need for transdimensional inference where the number of model components $N$ is treated as a discrete parameter to be sampled.
We use the \tbilby~code package \citep{tbilby_paper}, which approaches transdimensional sampling using \textit{ghost parameters}~\citep[see Ref.][for details]{tbilby_paper}.

\section{Glitch models}\label{sec:models}
We implement three different transdimensional glitch models in \bg, allowing the choice of model to suit the glitch morphology. The first is a physically motivated slow scattering model for common slow scattering glitches. The second is a sine-Gaussian wavelet that, similarly to \bw, can fit a broad range of morphologies. The third is a chirplet model that closely resembles a GW signal. 
Chirplets are useful for modeling the components of glitches that most resemble GW signals.
(For this study, we do not care about the quality of the glitch fit for its own sake; our only goal is to achieve unbiased inference of the binary parameters.)

\subsection{Slow scattering}

The induced strain due to scattering of light off of a surface may be written as ~\cite{Accadia:2010zzb, Soni:2023kqq}
\begin{equation}
    g(t) = A \sin\biggr{[}\frac{4\pi}{\lambda} \biggr{(}\delta x(t) + x_0 \biggr{)}\biggr{]},
\end{equation}
where $A$ is an amplitude term that depends on various factors, $\lambda=\unit[1064]{nm}$ is the wavelength of the carrier beam, and $\delta x(t) + x_0$ is the difference in the path length between the light scattering off the surface and the main beam, split into time varying and constant components.
Differences in the motion of the scatterer generate different types of glitches.
Slow scattering is produced by motion in the microseism band ($f\approx0.1 - 0.5~\text{Hz}$), and is often characterised by scattered light which bounces multiple times, increasing the path length such that
\begin{equation}
    x_0 + \delta x(t) \rightarrow n(x_0 + \delta x(t))
\end{equation}
for the $n$th arch. 

To model slow scattering, we assume that the motion $\delta x(t)$ is well described as a simple sinusoid~\cite{Udall:2022vkv, Tolley:2023umc}, such that
\begin{equation}
    \delta x(t) = \delta x_0 \sin\left(2 \pi f_\text{mod} (t - t_c)\right), 
\end{equation}
where $f_\text{mod}$ is the modulation frequency and $t_c$ is the time of the glitch.
Under the stationary phase approximation, the frequency of the $n$th scattering arch is
\begin{equation}
    f(t) = \biggr{|}\frac{4 \pi n f_\text{mod} \delta x_0}{\lambda} \cos(2 \pi f_\text{mod} (t-t_c)) \biggr{|} . 
\end{equation}
Denoting the peak frequency of the $n^\text{th}$ arch as $f_{h, n}$, we write the induced strain for the $n^\text{th}$ arch as
\begin{equation}
    g_n(t) = A_n \sin\biggr{[}\frac{f_{h,n}}{f_\text{mod}}\sin(2 \pi f_\text{mod}(t-t_c)) + \phi_n\biggr{]} .
\end{equation}
The total strain is 
\begin{align}
        g(t) =& \sum_{n=0}^N g_n(t).
\end{align}
Here we assume $A_n$ and $\phi_n$ are uncorrelated, while $f_{h, n} = n\delta f + f_{h, 0}$.
The variable $\delta f$ represents the spacing in peak frequencies between adjacent arches.
Following the notation of~\citet{Udall:2022vkv}, we index scattering arches from the lowest observable arch (usually $16 \lessapprox f_{harm}/{\rm Hz} \lessapprox 24$), designating this arch as $k=0$.

\begin{figure}[h]
    \centering
    \includegraphics[width=0.99\linewidth]{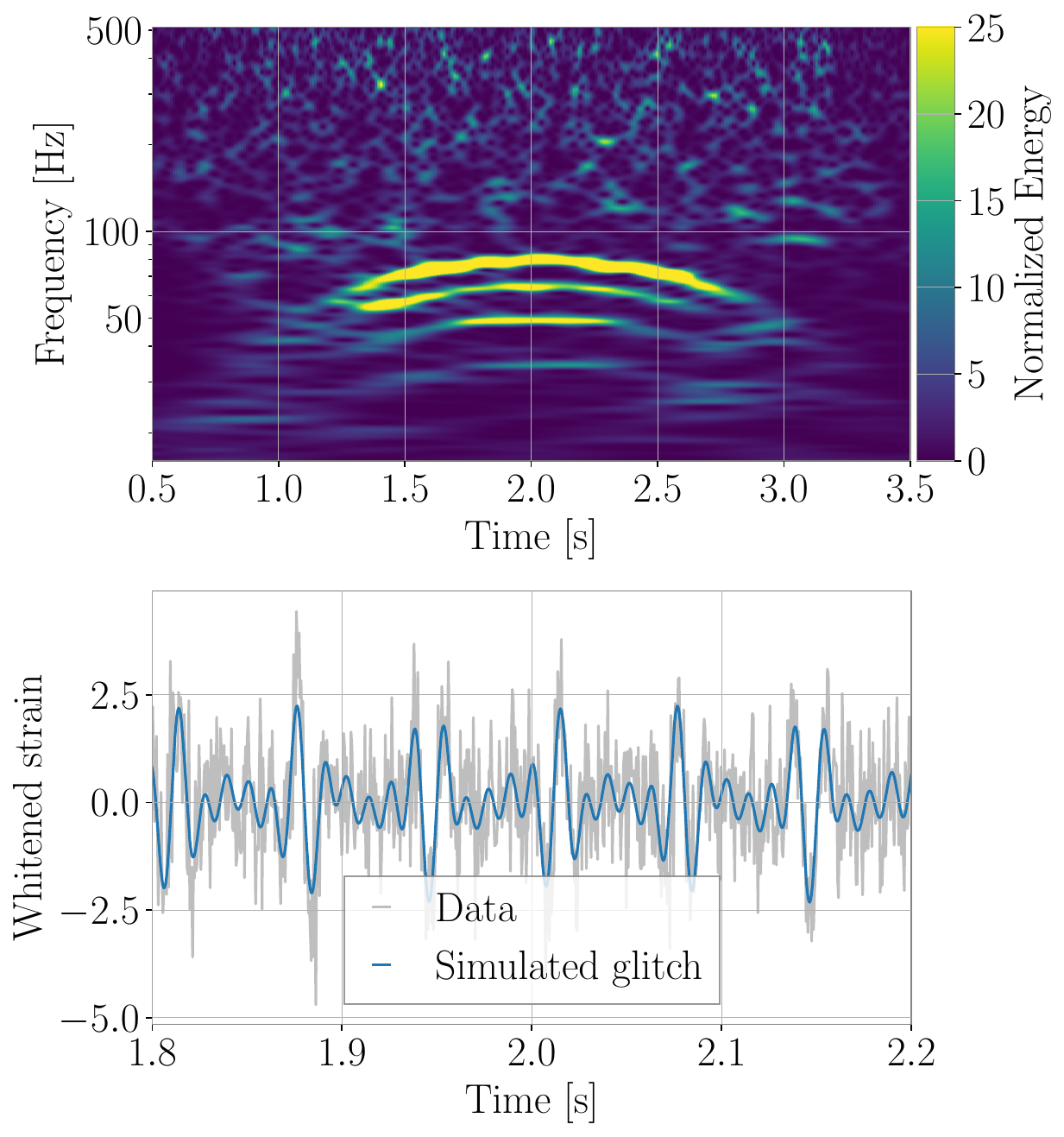}
    \caption{A simulated example of a slow scattering glitch, with five slow scattering arches simulated into coloured Gaussian noise. Top: spectrogram of a slow scattering glitch. Bottom: whitened time-domain reconstruction of the data (grey) and simulated glitch model (blue).}
    \label{fig:example_slow_scatter}
\end{figure}

The top panel of of Fig. \ref{fig:example_slow_scatter} shows an spectrogram of a slow scattering glitch from our model.
The lower panel shows the same glitch in the time domain.

\subsection{Wavelets}

GW detector noise features many glitches for which we do not have physical models.
Following Ref.~\cite{Cornish:2014kda}, we use a superposition of sine-Gaussian wavelets as a flexible model for unmodeled glitches.
A single sine-Gaussian wavelet can be written as
\begin{align}
    g(t) = A {\rm exp}\left[-\frac{(t-t_0)^2}{\tau^2} \right]\cos\left[2\pi f_0(t-t_0)+\phi\right],
\end{align}
where $A$ is the amplitude, $f_0$ is the central frequency, $t_0$ is the central time, $\tau$ is the damping time, and $\phi$ is the phase.
The damping time can be written as $\tau=Q/(2\pi f_0)$ with quality factor $Q$. 

When it comes time to fit the data with multiple wavelets, it can be challenging to assign labels to two wavelets with similar properties.
Ref.~\cite{tbilby_paper} pointed out that it can be useful to useful to label component functions according to their signal-to-noise ratio (SNR).
For this purpose, we note that the SNR of a wavelet can be approximated as

\begin{equation}\label{eq:SNR_to_A_wavelet}
    \text{SNR} \approx
    \frac{A\sqrt{Q}}{\sqrt{2\sqrt{2\pi}f_0P(f_0)}}.
\end{equation}

\begin{figure}[h]
    \centering
    \includegraphics[width=0.99\linewidth]{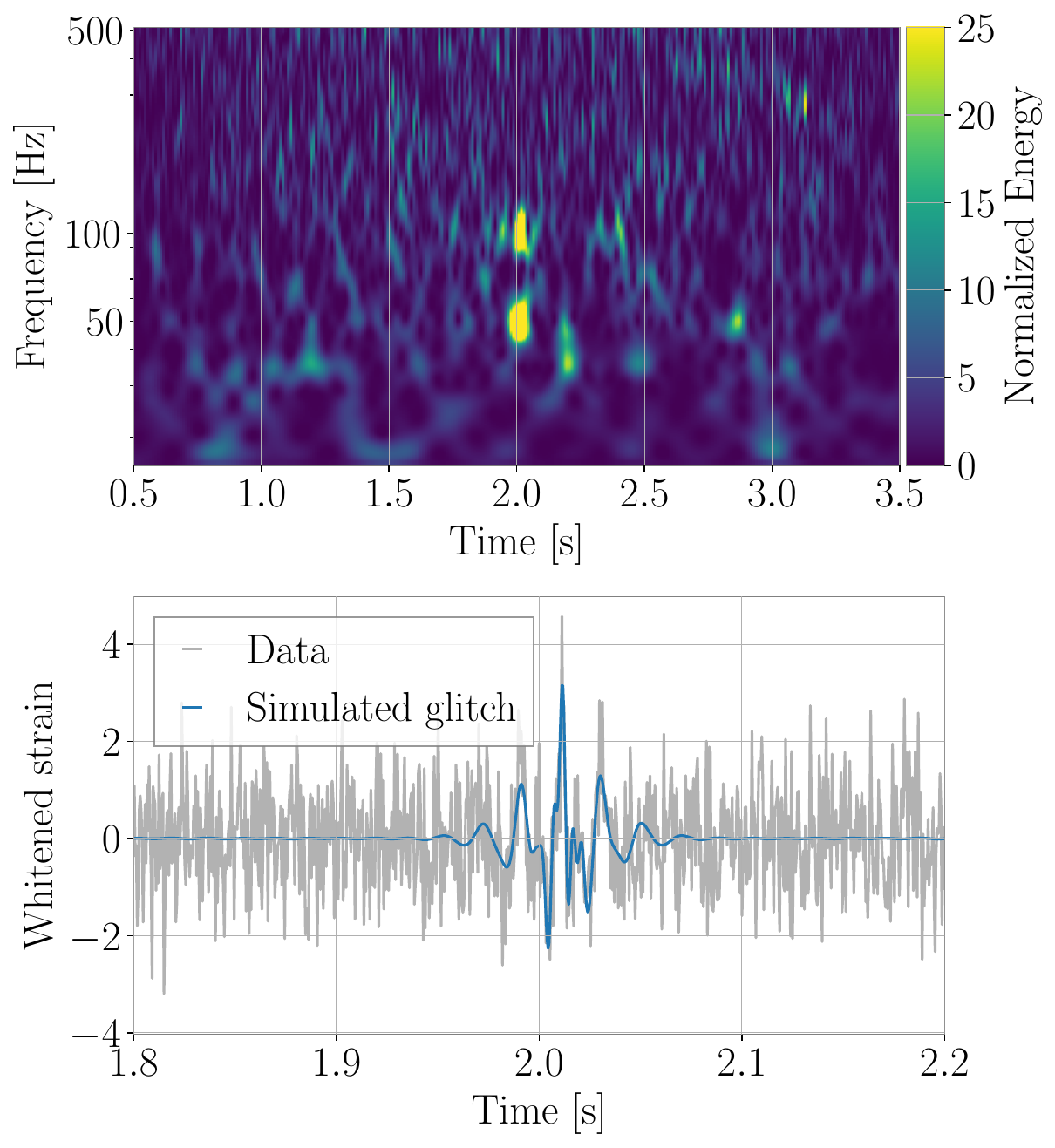}
    \caption{
    As Fig.~\ref{fig:example_slow_scatter} for a simulated wavelet glitch with a superposition of three wavelets simulated into coloured Gaussian noise at \unit[50]{Hz}, \unit[100]{Hz} and \unit[200]{Hz}.}
    \label{fig:example_wavelet}
\end{figure}

We use a superposition of $N$ wavelets to model a glitch
\begin{equation}
    g(t|N, \gamma_w) = \sum^N_{n=0} g(t|\gamma_{w,n}), 
\end{equation}
where $\gamma_{w, n}=\{A_n, f_{0, n}, t_{0, n}, \tau_n, \phi_n\}$.
Fig.~\ref{fig:example_wavelet} shows an example of a wavelet glitch from our model in a spectrogram (top plot) and the time domain reconstruction (bottom plot). 

\subsection{Chirplets}

\begin{figure}
    \centering
    \includegraphics[width=0.99\linewidth]{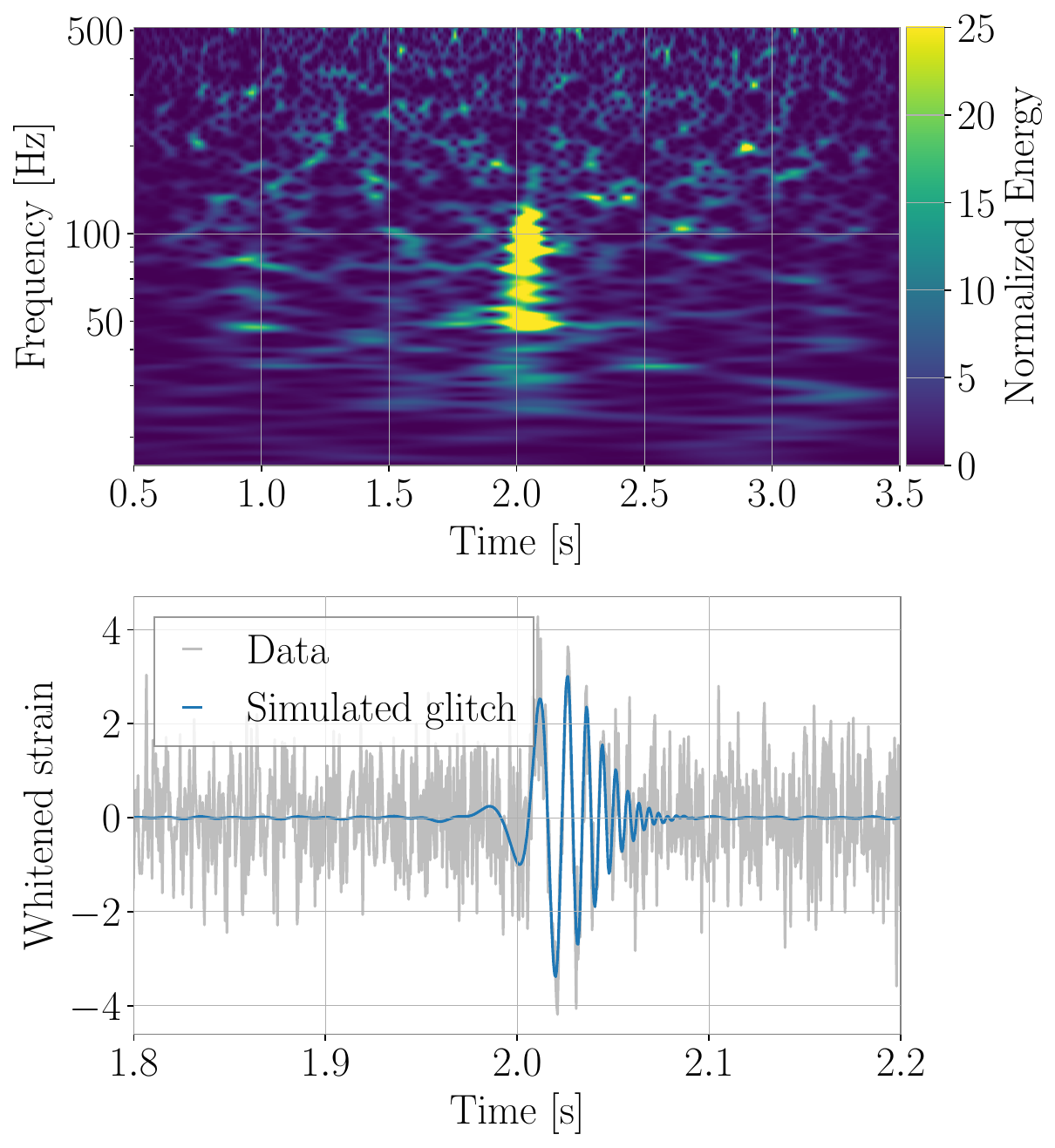}
    \caption{As Fig.~\ref{fig:example_slow_scatter} for a simulated chirplet glitch, with one chirplet simulated into coloured Gaussian noise.}
    \label{fig:example_chirplet_plot}
\end{figure}

A chirplet is a generalisation of a wavelet with an additional sixth parameter, the chirp rate $\beta$, which allows the frequency to evolve over time. 
\citet{Millhouse:2018dgi} and~\citet{Chatziioannou:2017ixj} studied the use of chirplets for GW inference, including both joint glitch-GW inference and neutron star post-merger signals, finding that while they can provide improved matches of glitches and of GW signals, the increased dimensionality actually degrades the overall signal reconstruction, penalising their use in practical glitch mitigation.
However, since chirplets are good matches for CBC signals, an artificial ``glitch'' composed of chirplets can reliably bias GW inference, and hence be used as an effective test case for our mitigation methods.
We define the chirplet model to be \cite{Mann_1995}
\begin{equation}
    g(t|\beta ,\gamma_{w,0}) = A e^{-\frac{(t-t_{0})^2}{\tau^2}}e^{2\pi i f_{0}(t-t_{0})}e^{\frac{i\beta(t-t_{0})^2}{2}}e^{i\phi},
\end{equation}
and hence our transdimensional model becomes
\begin{equation}
    g(t|N, \gamma_c) = \sum^N_{n=0} g(t|\beta_n, \gamma_{w,n}).
\end{equation}
Fig.~\ref{fig:example_chirplet_plot} shows an example of chirplet glitch in a spectrogram and as a time domain reconstruction.
There is not a simple approximation for the SNR of a chirplet, and so we calculate this manually for labelling purposes.

\section{Simulations}\label{sec:valid}
In this Section, we test and validate our \bg~pipeline for our different models using simulations. In all our simulations, we use \unit[4]{s} of data, a sampling frequency of \unit[2048]{Hz} and the frequency band of our data is \unit[16 - 896]{Hz}. We marginalise the luminosity distance in all our joint inference to reduce the dimensionality and decrease the computational cost. 

\subsection{Validation test for slow scattering}\label{sec:slow-scattering-validation-test}

We validate the slow scattering model by performing 145 binary black hole and and slow scattering simulations into coloured Gaussian noise in both LIGO Hanford and Livingston observatories. We recover the simulated parameters using simultaneous inference. Configurations for the signal model and slow scattering simulations are drawn from their respective priors, listed in Table \ref{tab:bbh_priors} and Table \ref{tab:slow_scattering_priors}, respectively. For both simulation and recovery of the signal model we use the waveform approximant \imrphenomXPHM~\cite{Pratten_2021}, which includes precession effects and higher order modes. Sampling is performed with \dynesty \cite{Speagle:2019ivv} using 2000 live points. A joint inference run with slow scattering typically increases run times by 50\% or less with respect to comparable signal-only analysis. 

\begin{figure}[htbp]
    \centering
    
    \begin{subfigure}{\linewidth}
        \centering
        \includegraphics[width=0.99\linewidth]{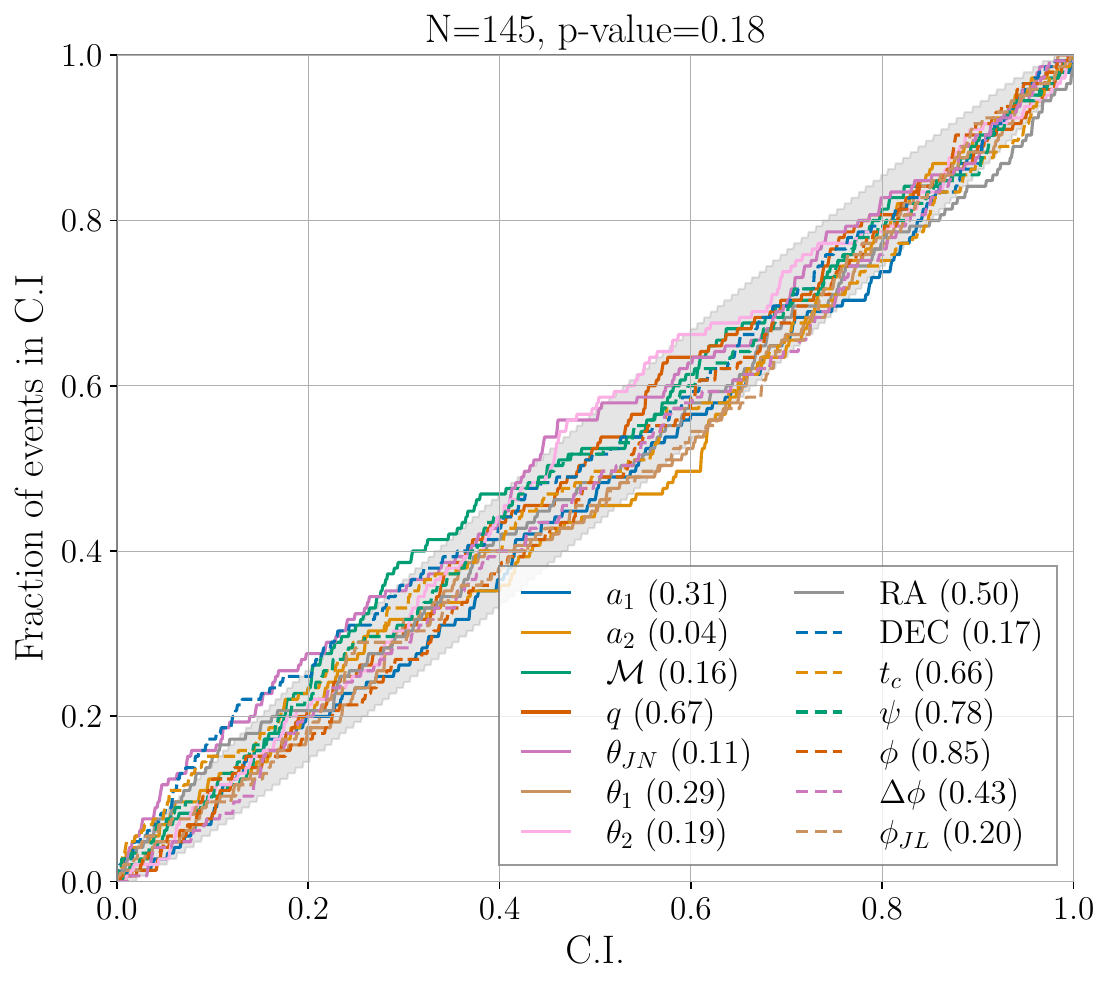}
        \label{fig:bbh_pp_plot}
    \end{subfigure}
    
    \begin{subfigure}{\linewidth}
        \centering
        \includegraphics[width=0.99\linewidth]{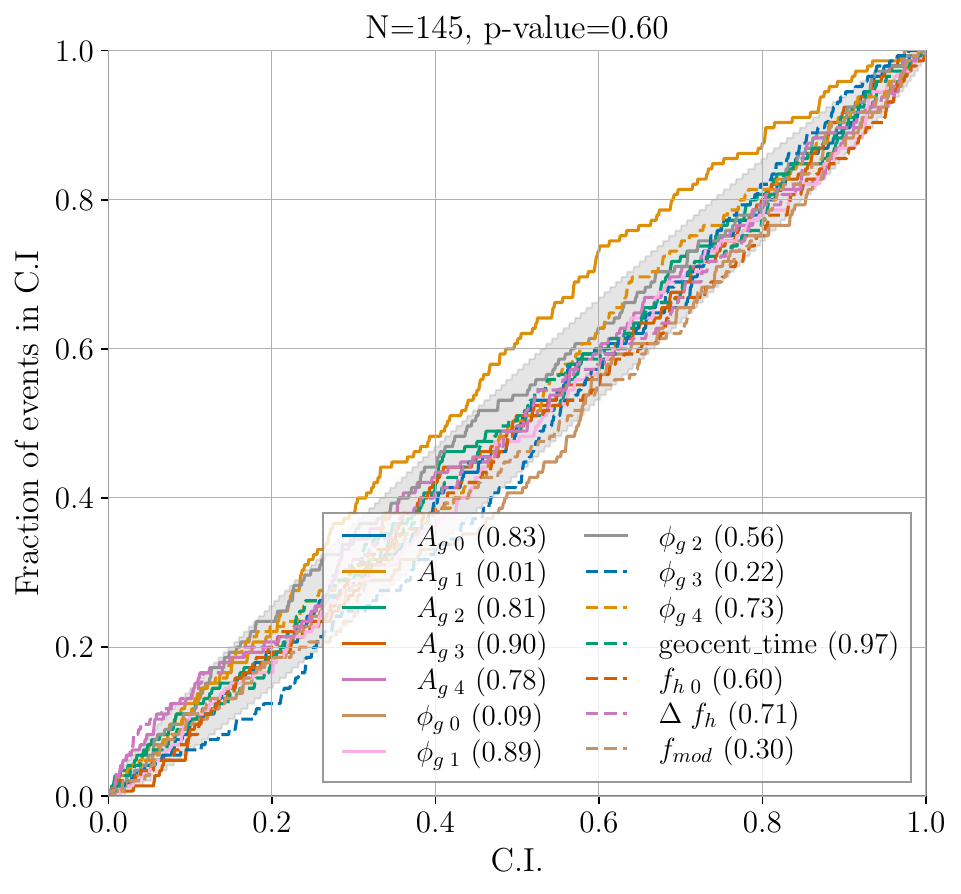}
        \label{fig:slow_scattering_pp_plot}
    \end{subfigure}
    
    \caption{Results for 145 simulations drawn from the priors in Table \ref{tab:bbh_priors} and \ref{tab:slow_scattering_priors}. The gray region covers the cumulative 90\% confidence interval. Each color curve tracks the cumulative fraction of events within the confidence interval on the horizontal axis for each parameter. The top plot shows the result for the binary black hole parameters and the bottom plot shows the result for the glitch parameters. The combined p-value for the signal-model and glitch parameters is 0.18 and 0.60, respectively, representing the probability that the inferred distribution is consistent with the expected distribution. The numbers in the legends are the individual p-values for each parameter. The luminosity distance parameter is not included as we marginalise over this parameter.}
    \label{fig:both_pp_plot}
\end{figure}

\begin{figure*}[htbp]
    \centering
    \includegraphics[width=0.99\linewidth]{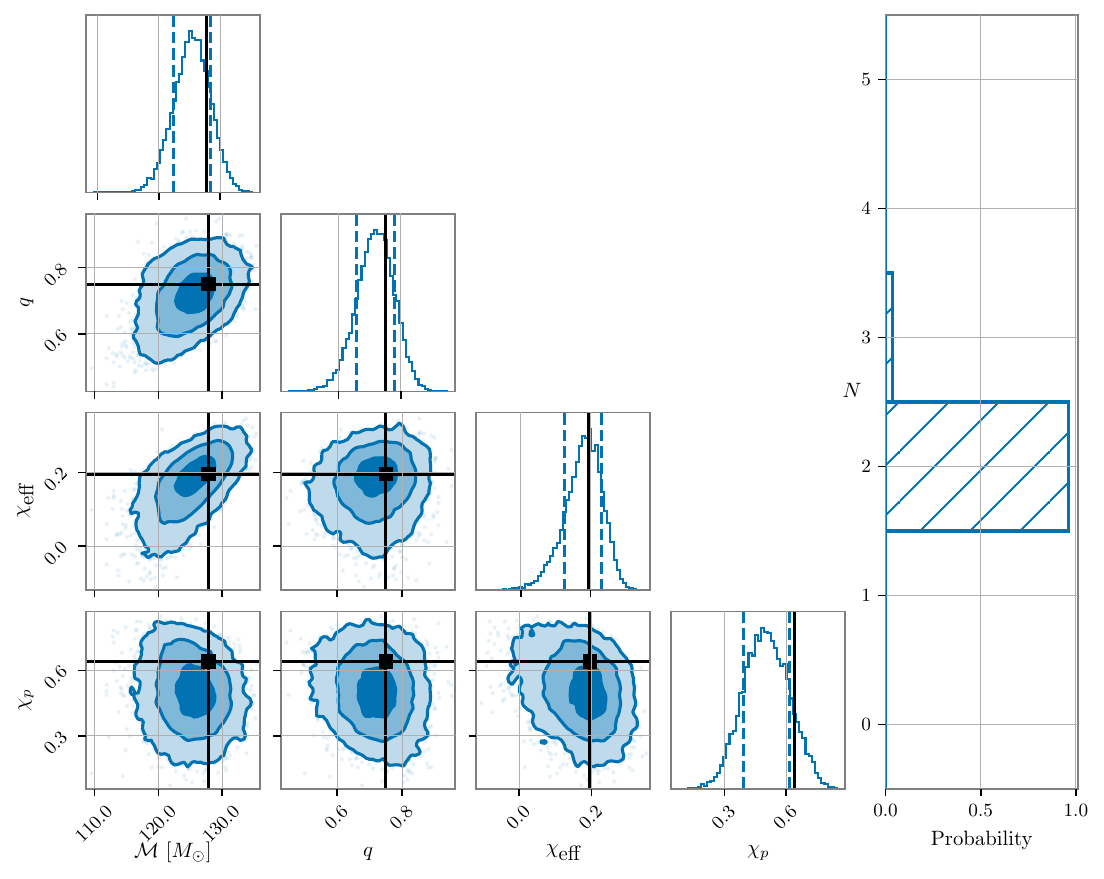}
    \caption{An example of signal+glitch simulation and recovery for a sample slow scattering glitch with $N=2$ arches. The right shows the posterior of the number of slow scattering arches, for which we recover the simulated value of $N=2$. The left shows posteriors of four signal parameters --- chirp mass $\mathcal{M}$, mass ratio $q$, effective inspiral spin parameter $\chi_\text{eff}$ and effective spin-precession parameter $\chi_p$ --- which recover the simulated value (black) within 2-$\sigma$. }
    \label{fig:slow_scatter_inject_posterior}
\end{figure*}

\begin{figure*}[htbp]
    \centering
    \includegraphics[width=0.99\linewidth]{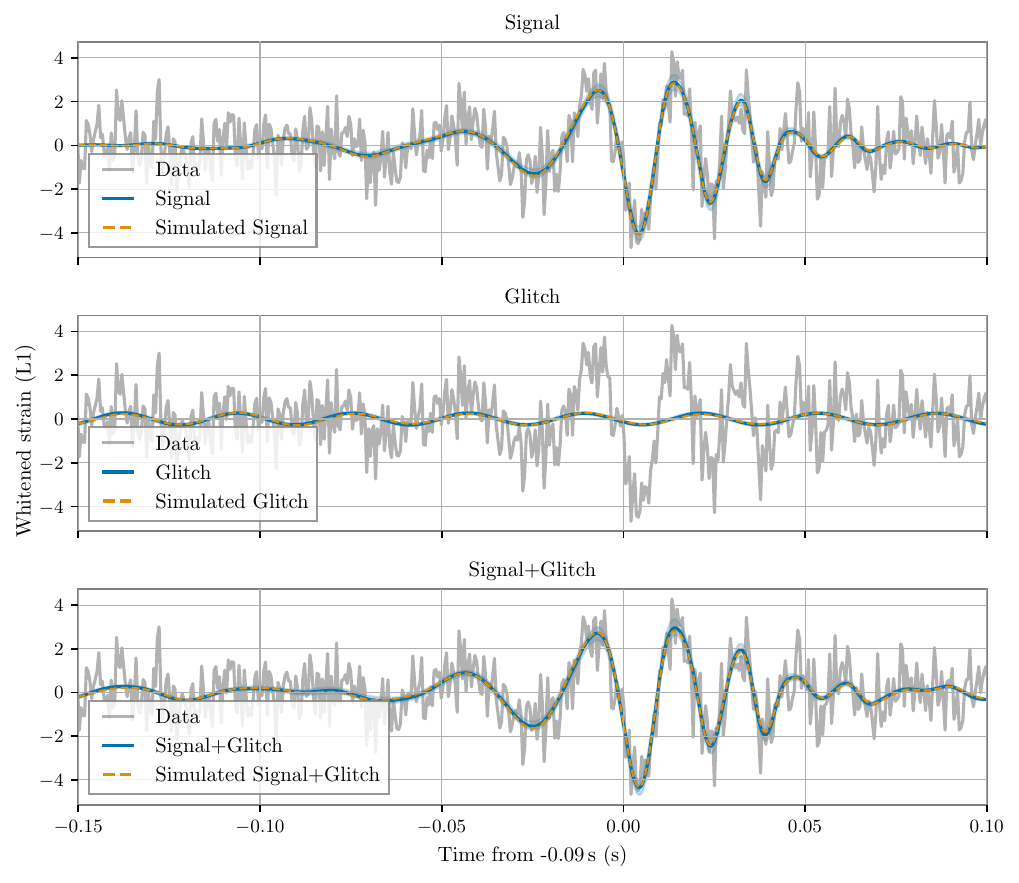}
    \caption{The time domain reconstruction of an example signal+glitch simulation with $N=2$ slow scattering arches, showing the signal (top), glitch (middle), and signal+glitch (bottom) from joint inference. The reconstruction for all three cases agree with their simulations within 90\% credible interval.}
    \label{fig:slow_scatter_inject_reconstruction}
\end{figure*}

In Fig.~\ref{fig:both_pp_plot} we show the probability-probability (P-P) plot for our signal-model (top panel) and for our glitch (bottom panel) parameters. Additionally, we show the p-value from a KS test for each individual parameter and the combined value. These plots show that \bg{} produces well-behaved credible intervals.\footnote{The transdimensional sampler employs ghost parameters, which recover the prior when deactivated. Since these uniform priors contribute as noise and thus follow the diagonal of the P-P plot, we include all glitch parameters, both active and ghost. }  

We see that for all signal-model parameters (top panel), the results are consistent with the expected diagonal. For the slow scattering parameters, almost all parameters fall within the 90\% credible interval, except for $A_{g,1}$ which has a p-value of 0.01. 
However, as noted above, our priority is the inference of the signal parameters, and so the convergence of our sampling is adequate for our goal.

We show an example of a typical signal+glitch simulation and recovery in Fig.~\ref{fig:slow_scatter_inject_posterior} and \ref{fig:slow_scatter_inject_reconstruction}. The right panel of Fig.~\ref{fig:slow_scatter_inject_posterior} shows the $N$ posterior is consistent with the simulated $N=2$ glitch. The left panel shows the posterior of four signal parameters chirp mass $\mathcal{M}$, mass ratio $q$, effective inspiral spin parameter $\chi_{\text{eff}}$ and effective spin-precession parameter $\chi_p$, which are consistent the simulated value. The bottom panel of Fig.~\ref{fig:slow_scatter_inject_reconstruction} shows the time domain reconstruction of the signal+glitch matches the simulated signal+glitch within 90\% credible interval. 

\subsection{Unbiasing astrophysical parameters with chirplets}

It is computationally intensive to sample both the signal and wavelet parameters with uninformative priors and the \imrphenomXPHM waveform, taking up to five days with 16 CPU cores. Running a sufficient number of simulations required to do a comprehensive validation test would be too computationally expensive for a conventional sampler such as \dynesty. Hence, we instead show a case in which the glitch configuration is engineered to bias the astrophysical inference, and demonstrate that the marginalisation method with wavelet/chirplet can reduce these biases. However, one can perform the validation test for the wavelet model with uninformative priors given enough computational resources. 

We simulate the signal from a non-spinning binary black hole and coloured Gaussian noise in both the Hanford and Livingston detectors, and additionally simulate an overlapping chirplet glitch in the Livingston detector.
Table \ref{tab:chirplet_simulation_values} shows the parameters used to simulate both the signal and the glitch. We perform joint inference using \dynesty~with 3000 live points. To simplify the problem and reduce computational cost, we assume a non-spinning system, choosing not to sample the spin parameters $\{a_1, a_2, \phi_{12}, \phi_{jl}, \theta_1, \theta_2\}$, and adopt the computationally fast \imrphenompv~waveform~\cite{Hannam:2013oca, Schmidt:2014iyl}.

\begin{table}[htbp]
    \centering
    \caption{simulated values for both signal-model and glitch-chirplet parameters. The simulated values of parameters $a_1=a_2=\psi = \Psi = \phi_{12} =\phi_{JL}=\theta_1=\theta_2=0$. }
    \label{tab:chirplet_simulation_values}
    \begin{tabular}{ll}
        \hline
        Parameter & simulated Value \\
        \hline
        $\mathcal{M}$ & $\unit[60]{M_\odot}$ \\
        q & 0.8 \\
        $d_L$ & \unit[4000]{Mpc} \\
        $t_{c, \text{GW}}$ & \unit[0]{s}  \\
        $\alpha$ & \unit[2.1]{rad}\\
        $\delta$ & \unit[-1.3]{rad}\\
        $\theta_{JN}$ & \unit[0.3]{rad}\\
        $N$ & 1 \\
        $\text{SNR}_0$ & 12 \\
        $Q_0$ &  12.7 \\
        $t_{\text{glitch}, 0}$ & \unit[0.02]{s}  \\
        $f_0$ & \unit[105]{Hz} \\
        $\phi_0$ & \unit[4.7]{rad}  \\
        $\beta_0$ & \unit[8000]{Hz/s} \\
        \bottomrule
    \end{tabular}
\end{table}

\begin{figure*}[htbp]
    \centering
    \includegraphics[width=0.99\linewidth]{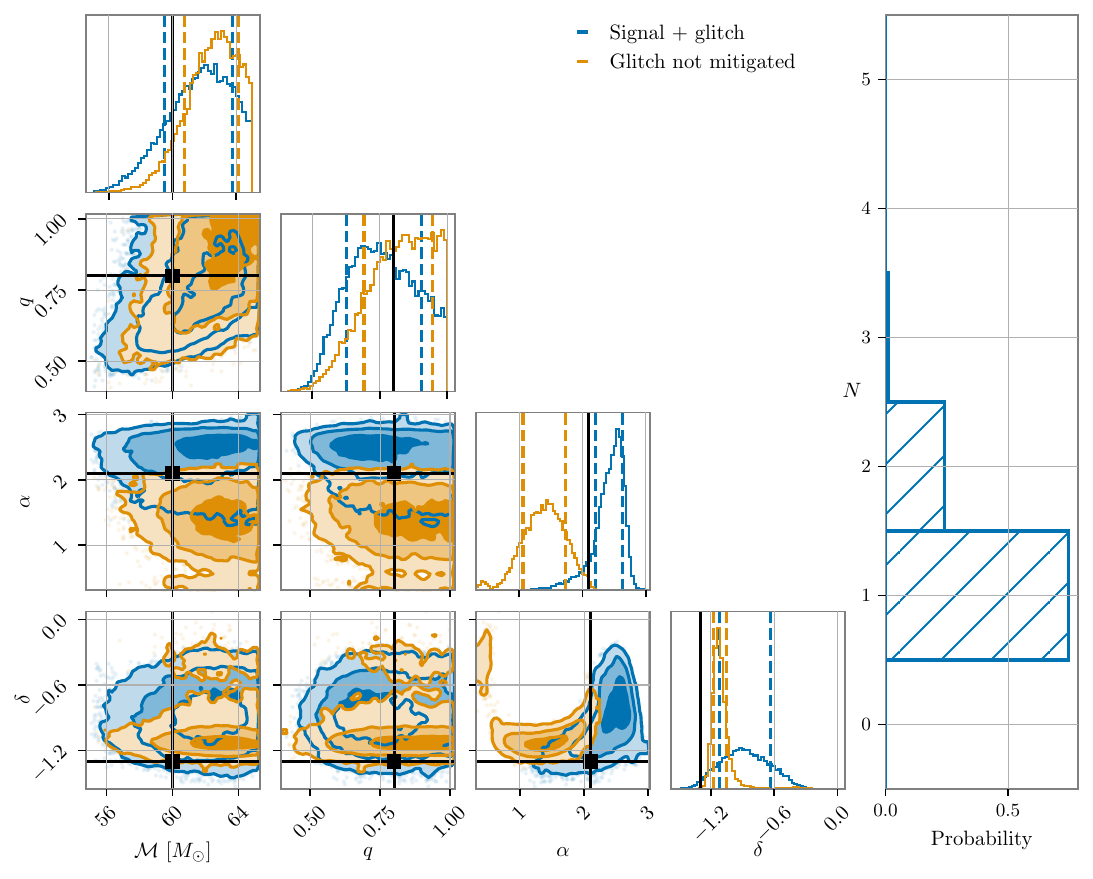}
    \caption{Results from an example of signal+glitch simulation and recovery for a non-spinning BBH and a chirplet model with $N=1$. 
    The corner plots shows posteriors for chirp mass $\mathcal{M}$, mass ratio $q$, right ascension $\alpha$ and declination $\delta$ with the simulated values shown in black (left). The posterior of the number of chirplets peaks at the simulated value (right).}
    \label{fig:chirplet_corner}
\end{figure*}

\begin{figure*}[htbp]
    \centering
    \includegraphics[width=0.99\linewidth]{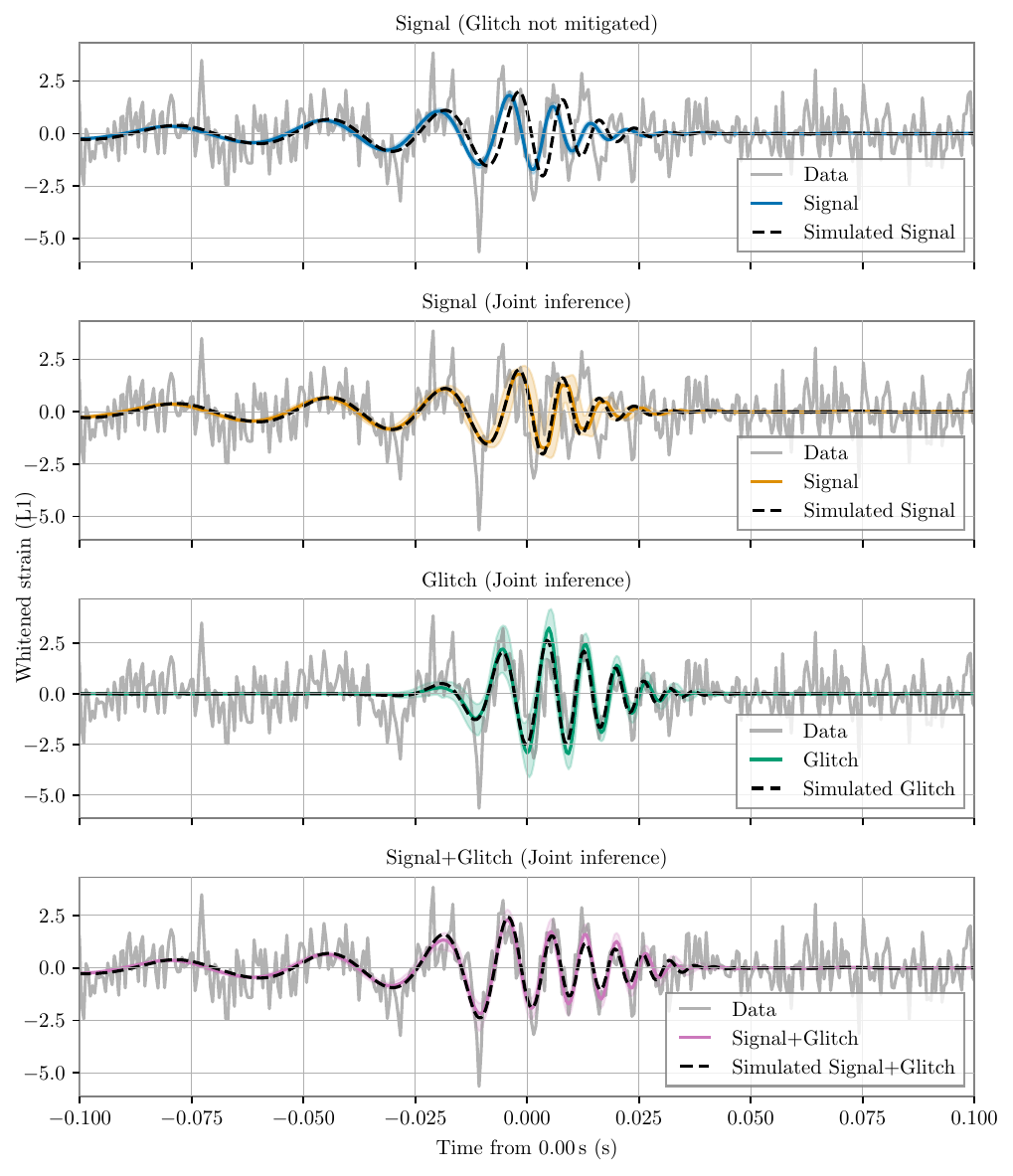}
    \caption{Results from an example of signal+glitch simulation and recovery for a non-spinning BBH and a chirplet model with $N=1$, showing time domain reconstructions for (top to bottom) the signal where the glitch is not mitigated, the individual signal and glitch components from joint inference respectively, and the signal+glitch from joint inference. The signal+glitch reconstruction (pink) is in better agreement with the data (grey) than the signal where the glitch was not mitigated (blue)}
    \label{fig:chirplet_reconstruction}
\end{figure*}

Table \ref{tab:chirplet_bbh_priors} lists the signal-model and glitch priors used in the run.
Following Ref.~\cite{tbilby_paper}, we order component functions according to their SNR.
The results are shown in Fig.~\ref{fig:chirplet_corner} and \ref{fig:chirplet_reconstruction}.
The right of Fig.~\ref{fig:chirplet_corner} shows that we recover the simulated number of chirplets $N=1$. The left of Fig.~\ref{fig:chirplet_corner} compares the posteriors for chirp mass, mass ratio, right ascension $\alpha$ and declination $\delta$ for analyses with and without glitch mitigation in blue and orange, respectively.
The posteriors from both cases are noticeably different, showing that the misspecified noise model is inadequate for reliable inference. We find that including the glitch model broadens the posteriors for the signal parameters when calculating the 90\% credible interval. Chirp mass, declination and zenith angle $\theta_{jn}$ posterior is 25\%, 286\% and 550\% wider when using glitch marginalisation, respectively. Mass ratio and phase posteriors are unchanged. The right ascension is 25\% narrower when using glitch mitigation.
In Fig.~\ref{fig:chirplet_reconstruction}, we plot the reconstructed time series for four cases: the signal+glitch, signal-only and glitch-only from our joint inference, and the signal where the glitch was not mitigated. The signal+glitch reconstruction obtained with the glitch model better agrees with the data, especially after $t=\unit[0]{s}$ compared to the reconstruction obtained where the glitch was not mitigated.

\section{Applications to real events}\label{sec:app}
In this section we revisit two GW events in the catalogue which have suspected data-quality issues. The first event is GW191109 \cite{GWTC-3_paper}, which has a scattered light glitch overlapping with the signal in LIGO Livingston, potentially biasing the effective spin inspiral parameter $\chi_\textrm{eff}$ \cite{Udall_2025}. The second event is GW200129, which has a glitch in LIGO Livingston \cite{GWTC-3_paper}, potentially biasing the effective spin-precession parameter $\chi_p$ \cite{Payne:2022spz}.

\subsection{Slow scattering - GW191109}\label{sec:gw191109}

The event GW191109 contains scattered light glitches in both LIGO Hanford and LIGO Livingston detectors \cite{GWTC-3_paper}.
However, the glitch in LIGO Hanford did not overlap with the GW signal, so its impact on parameter estimation is likely negligible \cite{Hourihane:2025vxc}.
However, the glitch in LIGO Livingston overlapped in time and frequency with the signal, which can lead to biased inference if the glitch modeling is not treated carefully~\cite{Udall_2025}. 

We perform joint inference with the transdimensional slow scattered light model using \dynesty~with 500 live points.
We model the GW signal using the \imrphenomXPHM waveform approximant. 
For our glitch priors, our prior on $N$ is discrete uniform between 1 to 9, all amplitude priors are uniform between $1\times 10^{-24}$ and $1\times 10^{-21}$, the base harmonic frequency prior is uniform between \unit[18]{Hz} and \unit[20]{Hz}, the modulation frequency prior is uniform between \unit[0.05]{Hz} and \unit[0.15]{Hz}, and $\delta f$ is uniform between \unit[5]{Hz} and \unit[8]{Hz}.

\begin{figure*}
    \centering
    \includegraphics[width=0.99\linewidth]{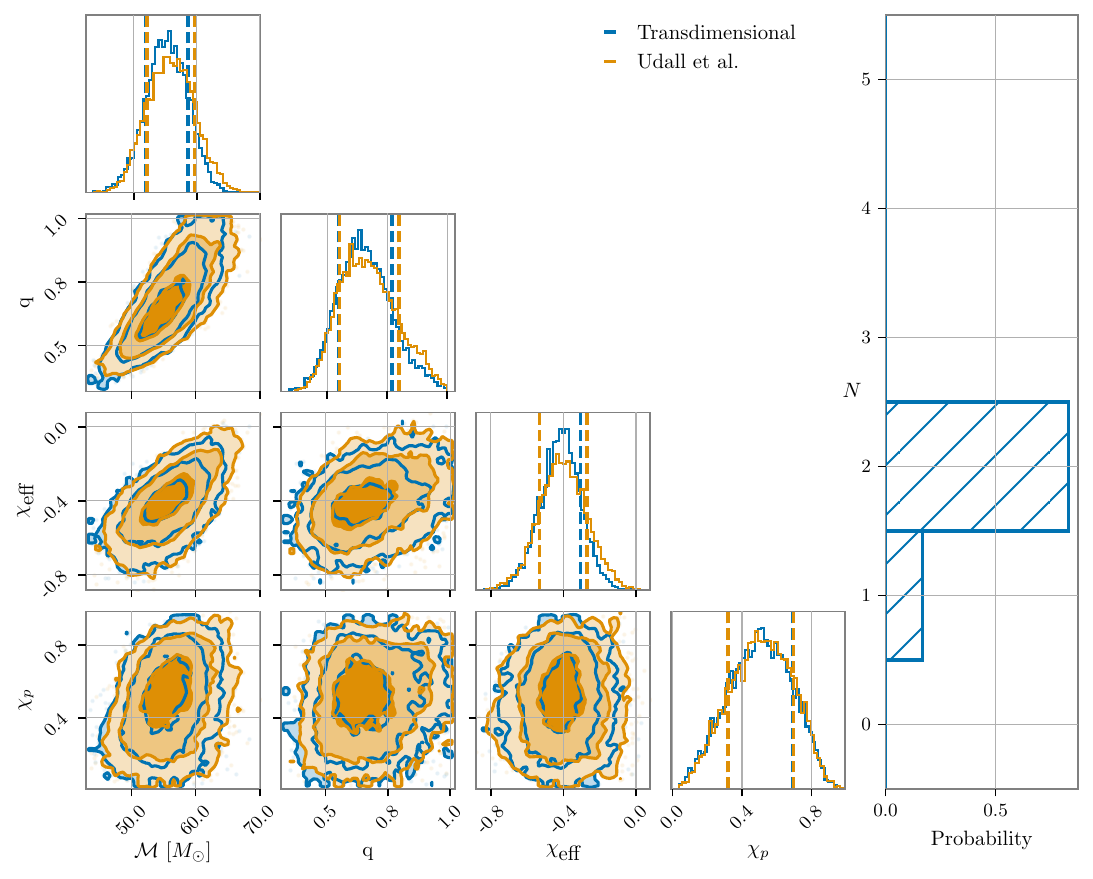}
    \caption{Results from the analysis of GW191109 with a transdimensional slow scattering model compared with \citet{Udall_2025} which assumed $N=5$. The right panel shows the posterior of the number of scattering arches for our analysis; notably, there is large posterior support for $N=2$ and no posterior support for $N>2$.
    The left panel shows the posteriors for chirp mass $\mathcal{M}$, mass ratio $q$, effective inspiral spin parameter $\chi_{\mathrm{eff}}$ and effective spin-precession parameter $\chi_p$.
    Our posterior (blue) is compared with the posteriors from \cite{Udall_2025} (orange), showing marginal change.}
    \label{fig:GW191109_corner_posterior}
\end{figure*}

\begin{figure*}
    \centering
    \includegraphics[width=0.99\linewidth]{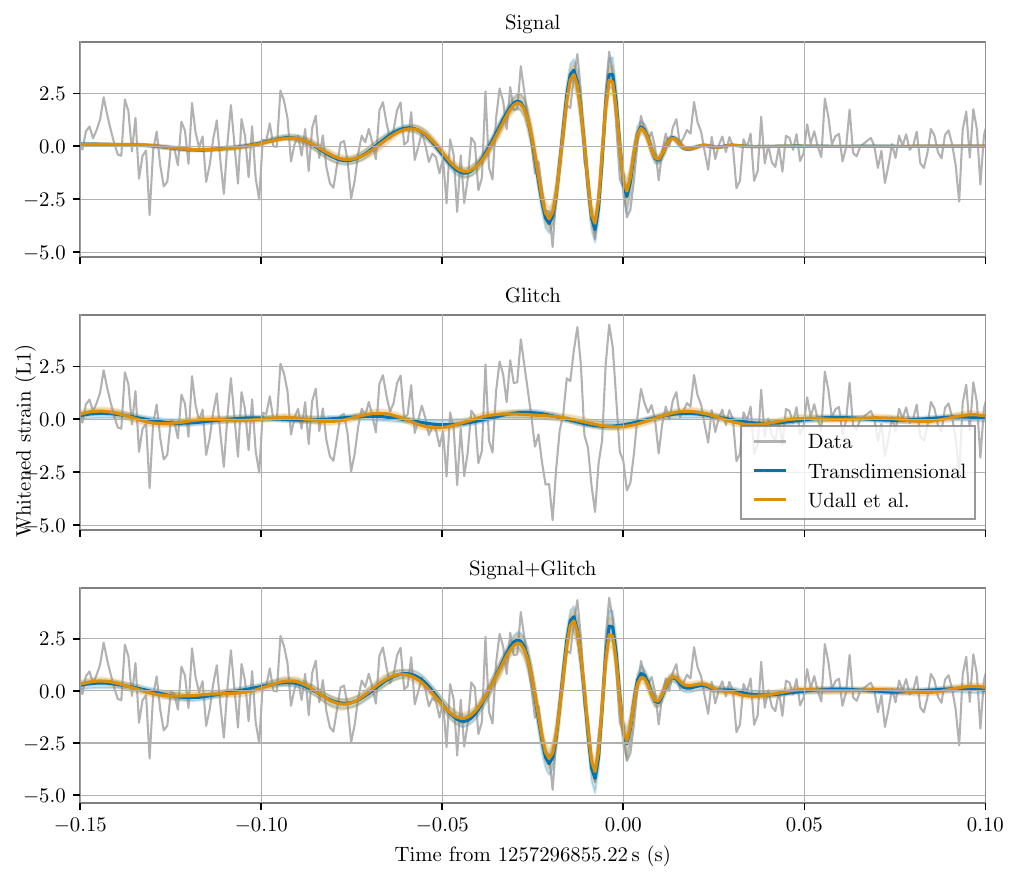}
    \caption{Whitened time domain reconstructions of GW191109 with transdimensional slow scattering model, showing the signal (top), glitch (middle), and signal+glitch (bottom) from joint inference. The reconstruction for our results (blue), the results in \cite{Udall_2025} (orange) and the data (grey) all agree with each other within statistical uncertainties.}
    \label{fig:GW191109_corner_reconstruction}
\end{figure*}

We show the posterior of the number of scattering arches $N$ in the right panel of Fig.~\ref{fig:GW191109_corner_posterior}, which has large support for $N=2$. We find no support for $N>2$, inconsistent with the the $N=5$ case used in ~\citet{Udall_2025}. In the left panel, we compare our posteriors for the intrinsic parameters to the $N=5$ case from \citet{Udall_2025}. The posteriors show marginal change compared to the results in \citet{Udall_2025}. 
Our posterior for $\chi_{\text{eff}}$ shows marginally more support for negative values. 
We show the time domain reconstruction of the recovered signal and glitch in the bottom panel of Fig.~\ref{fig:GW191109_corner_reconstruction} and find the data is within $90\%$ credible interval of the reconstruction.

\subsection{Wavelets - GW200129}\label{sec:gw200129}

The binary black hole merger GW200129 is an interesting event with potential evidence of spin-precession, but with data-quality issues in both LIGO Livingston and Virgo detectors~\cite{GWTC-3_paper}.
The LVK collaboration mitigated transient noise in the LIGO Livingston detector by using \textsc{gwsubtract}~\cite{Davis:2018yrz} to model correlations between auxiliary channels and the main strain channel, then subtracting the resulting correlated noise.
In GWTC-3~\cite{GWTC-3_paper} these glitch-subtracted data were used for parameter estimation with the \textsc{IMRPhenomXPHM} and \textsc{SEOBNRv4PHM} waveform approximants, which found strong evidence for spin-precession and no evidence for spin-precession respectively, indicating that waveform systematics play a significant role in interpretation of this event.
\citet{Hannam:2021pit} and ~\citet{Islam:2023zzj} then used these same glitch-subtracted data in parameter estimation with the \textsc{NRSur7dq4} waveform approximant, finding strong evidence for spin-precession. 
Subsequently,~\citet{Payne:2022spz} considered the impacts of differing glitch mitigation strategies on astrophysical inferences, using \bw{} inference --- including simultaneous inference with an aligned-spin model and wavelet-based glitch mitigation --- to show that different possible glitch realisations produced dramatically different evidence for spin-precession.
~\citet{Payne:2022spz} also showed that data in Virgo were inconsistent with data in the other two detectors, contraindicating the use of Virgo data in any analysis. 
Finally,~\citet{Macas:2023wiw} used machine learning methods informed by auxiliary models to produce new glitch-subtracted data and perform parameter estimation with \textsc{NRSur7dq4}, finding results similar to~\citet{Hannam:2021pit}.

We seek to extend the approach of~\citet{Payne:2022spz} by performing simultaneous inference of the BBH signal with \textsc{NRSur7dq4} and the glitch using a wavelet model. 
We perform this analysis using \dynesty~with 3000 live points, and restrict ourselves to the data from LIGO Livingston and LIGO Hanford in accordance with the findings of~\citet{Payne:2022spz}. As GW200129 is strongly affected by waveform systematics, we use both waveforms \imrphenomXPHM and \textsc{NRSur7dq4}. A joint inference run with \imrphenomXPHM and a wavelet/chirplet model typically takes \unit[1-5]{days} with 16 CPU cores. However, the computational cost can be reduced by constraining the prior. Using the constrained priors shown in Table~\ref{tab:GW200129_priors}, the inference takes only \unit[1-2]{days}. \citet{Payne:2022spz} found that the spin-precession evidence is concentrated in the \unit[20-50]{Hz} band, coinciding with the \unit[30-60]{Hz} glitch found in \citet{Davis:2022ird}. Although the glitch is known to exist between \unit[30-60]{Hz}, we choose to focus on the frequency range that contains the spin-precession evidence, and thus set the central frequency-wavelet priors to be \unit[20-50]{Hz}. \textsc{NRSur7dq4} has a valid total mass range of $M>60{M_\odot}$, hence we add this additional constraint in our \textsc{NRSur7dq4} run \cite{Varma:2019csw}. 

\begin{figure*}[htbp]
    \centering
    \includegraphics[width=0.99\linewidth]{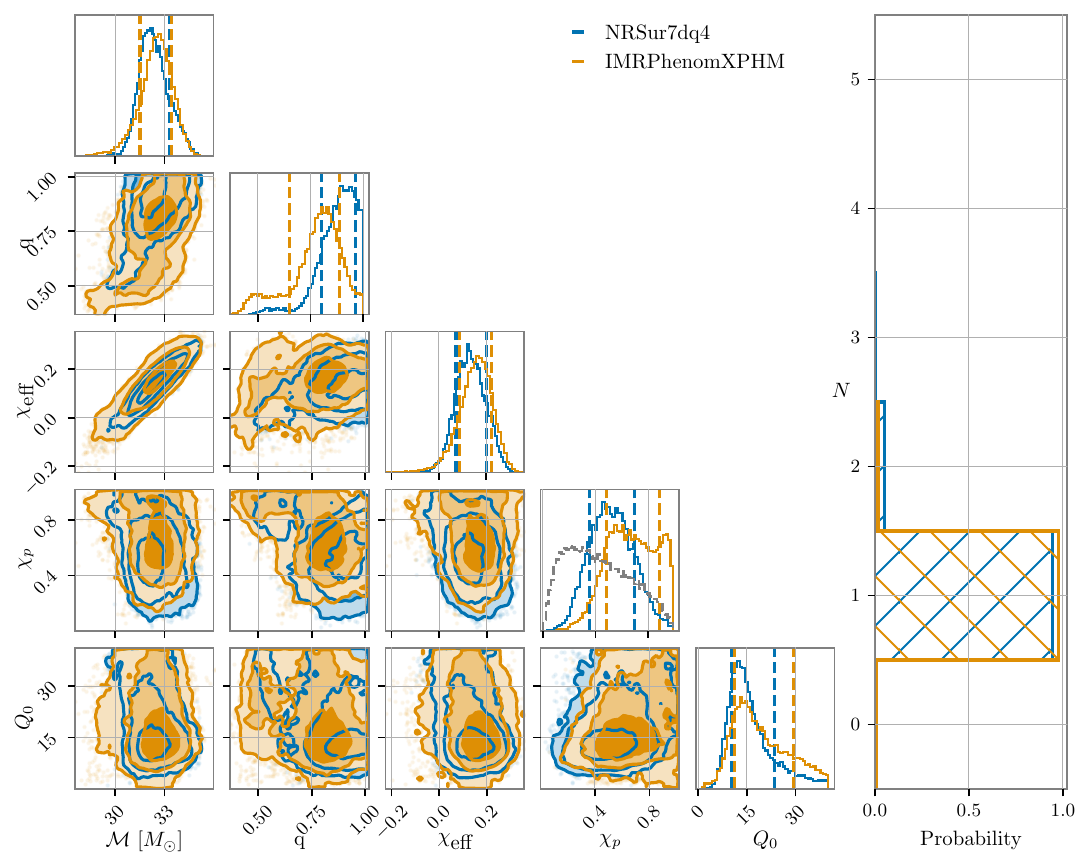}
    \caption{Posterior for chirp mass $\mathcal{M}$, mass ratio $q$, effective inspiral spin parameter $\chi_{\text{eff}}$, effective spin-precession parameter $\chi_p$ and the wavelet quality factor $Q_0$ for the GW200129 joint inference (left). We overlay the $\chi_p$ prior in grey in the $\chi_p$ histogram for reference. The two waveforms \textsc{NRSur7dq4} (blue) and \imrphenomXPHM (orange) show substantial difference in the $q$ and $\chi_p$ posteriors, with \textsc{NRSur7dq4} having prior-like spin-precession and \imrphenomXPHM supporting a maximally precessing system. Despite significant differences, the posteriors of both waveforms are not entirely disjoint. We find that $Q_0$ substantially different between the two waveforms and is correlated with $q$ and $\chi_p$. Posterior for number of wavelets for the GW200219 joint inference for both waveforms (right). We find strong posterior support for one wavelet.}
    \label{fig:GW200129_corner}
\end{figure*}

\begin{figure*}[htbp]
    \centering
    \includegraphics[width=0.99\linewidth]{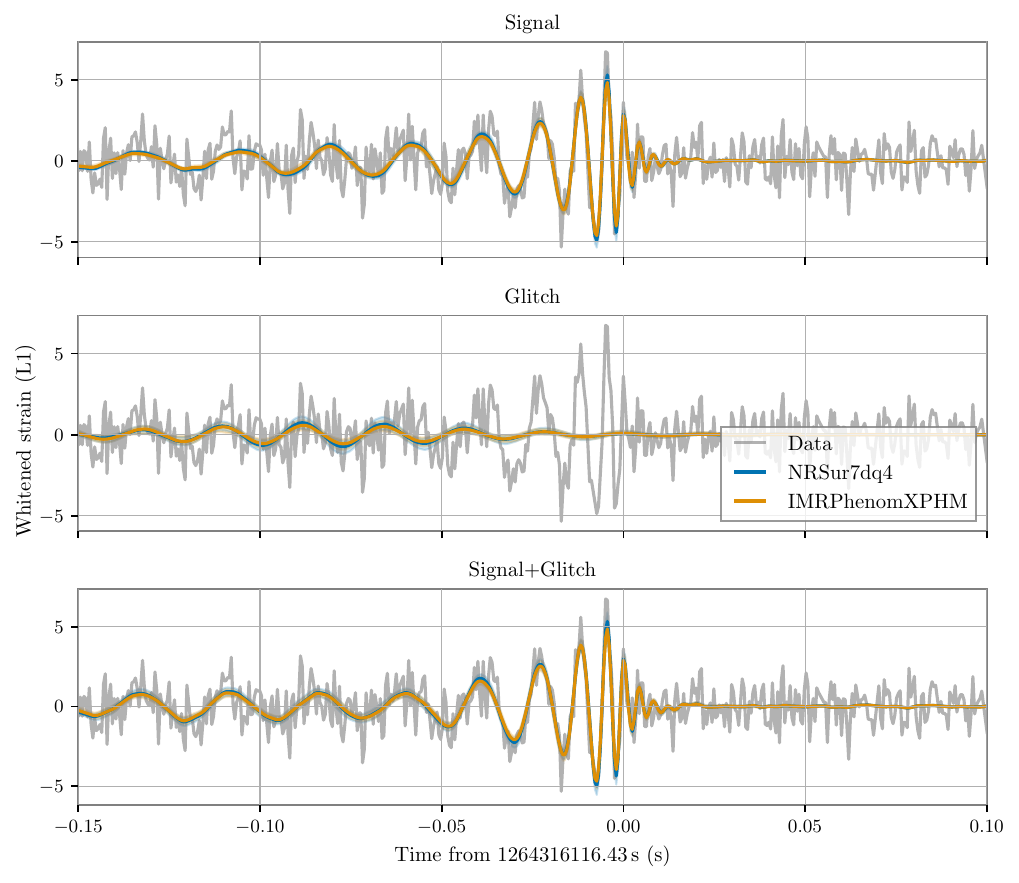}
    \caption{The time domain reconstruction of the GW200129 signal (top), glitch (middle), and signal+glitch (bottom) from joint inference using \textsc{NRSur7dq4}  (blue) and \imrphenomXPHM (orange). The signal+glitch reconstructions for both waveforms agree with the data within 90\% credible interval.}
    \label{fig:GW200129_reconstruction}
\end{figure*}

We compare the results from \imrphenomXPHM and \textsc{NRSur7dq4} in Fig.~\ref{fig:GW200129_corner} and \ref{fig:GW200129_reconstruction}.
In the right of Fig.~\ref{fig:GW200129_corner}, we see that the glitch in GW200129 strongly prefers one wavelet for both waveforms.
In the left of Fig.~\ref{fig:GW200129_corner}, we compare the intrinsic parameter posteriors between \imrphenomXPHM and \textsc{NRSur7dq4}, and found significant difference in $q$ and $\chi_p$.
The two waveforms prefer different astrophysical conclusions: where \textsc{NRSur7dq4} prefers a system with equal mass and weaker spin-precession with $\chi_p=0.52^{+0.29}_{-0.26}$, \imrphenomXPHM instead prefers a more unequal mass system and stronger spin-precession with $\chi_p=0.67^{+0.28}_{-0.30}$.
This large difference in posteriors suggest that this event is significantly affected by waveform systematics. We find negligible change in the wavelet posteriors between the two waveforms, apart from the quality factor $Q_0$ (included in Fig~\ref{fig:GW200129_corner}) and find that $Q_0$ is correlated with $q$ and $\chi_p$. In Fig.~\ref{fig:GW200129_reconstruction}, we show that the time domain reconstruction of the signal+glitch for \textsc{NRSur7dq4} (blue), \imrphenomXPHM (orange) and data (grey) agree with each other within 90\% credible interval. 

\section{Conclusion}\label{sec:conclusion}

We present \bg, a pipeline that allows for joint inference between a GW signal and a glitch.
We validate \bg~results with a slow scattering model, and then use a chirplet model to simulate a biased GW waveform and demonstrate that joint inference can reconstruct the glitch and show that modeling the glitch in conjunction with the signal model is needed to produce unbiased estimates of the astrophysical parameters. 
We then apply \bg{} to two real events.
We first reanalyse the event GW191109 which is affected by a slow scattering glitch in the LIGO Livingston detector.
We find that the posteriors are mostly similar, with the transdimensional model preferring a marginally more negative $\chieff$ than a non-transdimensional slow scattering model.
Next, we reanalyse the event GW200129, which is affected by a glitch in the LIGO Livingston detector, using joint inference with a transdimensional sine-Gaussian wavelet model. 
We find that \textsc{NRSur7dq4} and \imrphenomXPHM diverge in astrophysical interpretation, with the joint inference weakening the evidence of spin-precession in the former, and leaving strong evidence in the latter.
The astrophysical conclusions of our \textsc{NRSur7dq4} result are consistent with \citet{Payne:2022spz}, which performed \textsc{NRSur7dq4} analysis in conjunction with wavelet subtraction, but inconsistent with \citet{Hannam:2021pit}, which performed analysis with the same waveform but used \textsc{gwsubtract} subtraction. 

We view this paper as an important milestone in facilitating joint inference between signal and noise glitch in an easily usable Bilby-based pipeline.
Accordingly, we highlight potential future steps to further improve and expand \bg.
First, while \bg~currently only supports three glitch models, the number of glitch models can be greatly expanded.
We envision that researchers can easily write their own glitch model and include it into the pipeline to suit their own glitch mitigation needs. 
Existing glitch models in the literature that can be integrated into joint inference include \citet{glitschen, bondarescu_2023, malz_2025}.

Second, joint inference is computationally expensive. As \bg~is built on top of \bilby, it currently uses off-the-shelf samplers such as \dynesty~and \textsc{emcee} included in \bilby, which either do poorly with multimodality or high-dimensional parameter space (or both). 
We recommend that a sampler should be developed that supports transdimensional sampling and can handle the mulitmodality and scales well at high number of dimensions. 

In the process of finalising the paper, we became aware of the paper released in the LVK collaboration review by \citet{Hoy2026}, which similarly performs joint inference using the \textsc{antiglitch} model \cite{bondarescu_2023}. \citet{Hoy2026} reanalysed GW200129 and find strong evidence for spin-precession when using \textsc{NRSur7dq4} waveform and their own \textsc{antiglitch} model.
Although these results are opposite to our own \textsc{NRSur7dq4} results, this disagreement is unsurprising, since different glitch models are employed and, as this work and~\citet{Payne:2022spz} have shown, the inference of spin-precession in this event is highly sensitive to both glitch and waveform systematics. 

\acknowledgements

We thank Nir Guttman and Hui Tong for helpful discussions in using \tbilby. 
We thank Colm Talbot for helpful discussions on how to implement distance marginalisation with the joint likelihood.
We thank Katerina Chatziioannou for helpful discussion about the interpretation of GW200129. 
We thank Sophie Bini for their feedback during LVK collaboration internal review. 
We acknowledge support from the Australian Research
Council (ARC) Centres of Excellence CE170100004 and CE230100016, as well as ARC
LE210100002, and ARC DP230103088. S.Y.C receives support from the Australian Government Research
Training Program. 
RU was supported by the NSERC Alliance program.
This work was supported by a grant from the Simons Foundation International [SFI-MPS-SSRFA-00023625, DD].
This material is
based upon work supported by NSF’s LIGO Laboratory
which is a major facility fully funded by the National
Science Foundation. The authors are grateful for computational resources provided by the LIGO Laboratory
and supported by National Science Foundation Grants
PHY-0757058 and PHY-0823459.

This research has made use of data or software obtained from the Gravitational Wave Open Science Center (gw-openscience.org), a service of LIGO Laboratory,
the LIGO Scientific Collaboration, the Virgo Collaboration, and KAGRA. LIGO Laboratory and Advanced
LIGO are funded by the United States National Science Foundation (NSF) as well as the Science and Technology Facilities Council (STFC) of the United Kingdom, the Max-Planck-Society (MPS), and the State of
Niedersachsen/Germany for support of the construction
of Advanced LIGO and construction and operation of
the GEO600 detector. Additional support for Advanced
LIGO was provided by the Australian Research Council.
Virgo is funded, through the European Gravitational
Observatory (EGO), by the French Centre National
de Recherche Scientifique (CNRS), the Italian Istituto
Nazionale di Fisica Nucleare (INFN) and the Dutch
Nikhef, with contributions by institutions from Belgium,
Germany, Greece, Hungary, Ireland, Japan, Monaco,
Poland, Portugal, Spain. The construction and operation of KAGRA are funded by Ministry of Education,
Culture, Sports, Science and Technology (MEXT), and
Japan Society for the Promotion of Science (JSPS), National Research Foundation (NRF) and Ministry of Science and ICT (MSIT) in Korea, Academia Sinica (AS)
and the Ministry of Science and Technology (MoST) in
Taiwan.

\appendix

\section{Distance marginalisation for joint likelihoods}\label{sec:distance-marg}

To accelerate parameter inference, it is beneficial to analytically marginalise over parameters when possible.
One such parameter is luminosity distance, which may be marginalised over by use of a precomputed coefficients table~\cite{bilby_paper, Thrane:2018qnx,  Singer:2015ema, Udall:2025flo}.
This marginalisation uses the insight that gravitational waveforms scale inversely with distance, such that

\begin{equation}
    h(\theta, d_L) = h(\theta, d_0) \biggr{(}\frac{d_0}{d_L}\biggr{)}.
\end{equation}

Adopting the notation of~\citet{Thrane:2018qnx} one sees that the likelihood may be marginalised over distance

\begin{multline}
    \mathcal{L}_D (\kappa^2, \rho_{opt}) = \int \text{d} d_L \exp \biggr{\{} \biggr{(}\frac{d_0}{d_L}\biggr{)} \kappa^2(d_0) \\ - \frac{1}{2}\biggr{(}\frac{d_0}{d_L}\biggr{)}^2 \rho_{opt}(d_0)^2 \biggr{\}} \pi(d_L),
\end{multline}
where
\begin{equation}\label{eq:cbc_only_kappa}
    \kappa^2(d_L) = \langle d , h(\theta, d_L)\rangle,
\end{equation}
and
\begin{equation}
    \rho_{opt}^2 (d_L) =  \langle h(\theta, d_L) , h(\theta, d_L)\rangle, 
\end{equation}

such that 
\begin{equation}
    \ln \mathcal{L}_{\mathrm{marg}} = \ln \mathcal{Z}_N + \ln \mathcal{L}_D(\kappa^2, \rho_{\mathrm{opt}}).
\end{equation}
Since $\mathcal{L}_D (\kappa^2, \rho_{opt})$ is only a function of these two variables, a lookup table may be numerically computed for a given luminosity distance prior.

To extend this procedure to joint likelihoods, observe that the expansion of the joint likelihood only introduces one new set of inner products which depend on the gravitational waveform (and hence the luminosity distance):

\begin{equation}
    \kappa_g^2 = -\biggr{(}\frac{d_0}{d_L}\biggr{)}\langle g(\gamma) | h(\theta, d_0)\rangle.
\end{equation}

Since this term has the same luminosity distance dependence as $\kappa^2$ from before, Equation~\ref{eq:cbc_only_kappa} expands to the total factor

\begin{equation}
    \kappa_T^2 = \kappa^2 +\kappa_g^2 =  \langle d , h(\theta, d_L)\rangle - \langle g(\gamma) , h(\theta, d_L)\rangle.
\end{equation}

The full likelihood then has the expression
\begin{multline}
    \ln \mathcal{L}_{marg} = \ln \mathcal{Z}_N + \ln\mathcal{L}_D (\kappa_g^2, \rho_{opt}) \\ - \frac{1}{2} \langle g(\gamma) , g(\gamma)\rangle + \langle g(\gamma) , d\rangle.
\end{multline}
The same distance marginalisation tables work in this case, since the dependence on $\kappa_g^2$ is identical to the dependence on $\kappa^2$ in the signal-only case.

\section{Priors for joint inference}

In this section, we list all the signal and glitch priors used in all the joint inference runs in this paper. The signal and glitch priors for slow scattering validation test is listed in Table \ref{tab:bbh_priors} and \ref{tab:slow_scattering_priors}, respectively. The signal and glitch priors for the chirplet simulation is listed in Table \ref{tab:chirplet_bbh_priors}. The signal and glitch priors for GW200129 joint inference is listed in Table \ref{tab:GW200129_priors}.

\begin{table}[h]
    \centering
    \caption{Priors for the binary black hole parameters in our validation test of the slow scattering model. Priors for parameters $\alpha$, $\delta$, $\phi$, $\Psi$, $\phi_{12}$, $\theta_1$, $\theta_2$ ,$\phi_{JL}$ and $\theta_{JN}$ are the same as \cite{bilby_pipe_paper} and are not included in the table for brevity. The geocentric time $t_c$ prior is [\unit[-0.1]{s}, \unit[0.1]{s}] around the simulated value. \textit{UniformInComponents} refers to priors uniform in the component masses $m_1$ and $m_2$.}
    \label{tab:bbh_priors}
    \begin{tabular}{llll}
        \hline
        Parameter & Shape & Limits & Boundary \\
        \hline
        $a_1$ & Uniform & $[0, 0.88]$ & - \\
        $a_2$ & Uniform & $[0, 0.88]$ & - \\
        $\mathcal{M}$ & UniformInComponents & \unit[[12.3, 175]]{$M_\odot$} & - \\
        q & UniformInComponents & $[0.125, 1]$ & - \\
        $d_L (\text{Mpc})$ & PowerLaw ($\text{alpha}=2$) & \unit[[100, 2000]]{Mpc} & - \\
        \bottomrule
    \end{tabular}
\end{table}

\begin{table}[h]
    \centering
    \caption{Priors for slow scattering parameters in the validation of the slow scattering model.}
    \label{tab:slow_scattering_priors}
    \begin{tabular}{llll}
        \hline
        Parameter & Shape & Limits & Boundary \\
        \hline
        $N$ & DiscreteUniform & $[0, 5]$ & - \\
        $A_n$ & Uniform & $[1\times 10^{-24}, 1\times10^{-21}]$ & - \\
        $\phi_n$ & Uniform & \unit[[0, 2$\pi$]]{rad} & Periodic \\
        $f_{h, 0}$ & Uniform & \unit[[18, 22]]{Hz} & - \\
        $\delta f$ & Uniform & \unit[[10, 15]]{Hz} & - \\
        $f_{\text{mod}}$ & Uniform & \unit[[0.1, 0.3]]{Hz} & - \\
        \bottomrule
    \end{tabular}
\end{table}

\begin{table}[h]
    \centering
    \caption{Priors for both chirplet and signal-model parameters. Priors for parameters $\alpha$, $\delta$, $\psi$, $\Psi$ and $\theta_{JN}$ are the same as \cite{bilby_pipe_paper} and are not included in the table for brevity. \textit{DescendingOrder} are conditional priors that enforce the SNR parameters to be in descending order.}
    \label{tab:chirplet_bbh_priors}
    \begin{tabular}{llll}
        \hline
        Parameter & Shape & Limits & Boundary \\
        \hline
        $\mathcal{M}$ & UniformInComponent & \unit[[55, 65]]{$M_\odot$} & - \\
        $q$ & UniformInComponent & $[0.125, 1]$ & - \\
        $t_{c, \text{GW}}$ & Uniform & \unit[[-0.1, 0.1]]{s} & - \\
        $d_L$ & PowerLaw (alpha=0.2) & \unit[[3500, 4500]]{Mpc} & - \\
        $N$ & DiscreteUniform & [0, 5] & - \\
        $\text{SNR}_n$ & DescendingOrder & [0, 20] & - \\
        $Q_n$ & Uniform & [1, 20] & - \\
        $t_{\text{glitch}, n}$ & Uniform & \unit[[-0.1, 0.1]]{s} & - \\
        $f_n$ & Uniform & \unit[[16, 200]]{Hz} & - \\
        $\phi_n$ & Uniform & \unit[[0, $2\pi$]]{rad} & Periodic \\
        $\beta_n$ & Uniform & \unit[[0, 10000]]{Hz/s} & - \\
        \bottomrule
    \end{tabular}
\end{table}

\begin{table}[h]
    \centering
    \caption{Priors for both wavelet and binary black hole parameters used for GW200129. Priors for parameters $\alpha$, $\delta$, $\psi$, $\Psi$ and $\theta_{JN}$ are the same as Ref. \cite{bilby_pipe_paper} and are not included in the table for brevity. The time priors are relative to GPS time \unit[1264316116.4]{s}. }
    \label{tab:GW200129_priors}
    \begin{tabular}{llll}
        \hline
        Parameter & Shape & Limits & Boundary \\
        \hline
        $\mathcal{M}$ & UniformInComponent & \unit[[26, 38]]{$M_\odot$} & - \\
        $q$ & UniformInComponent & $[0.167, 1]$ & - \\
        $a_1$, $a_2$ & Uniform & [0, 0.99] & - \\
        $t_{c, \text{GW}}$ & Uniform & \unit[[-0.1, 0.1]]{s} & - \\
        $d_L$ & PowerLaw (alpha=0.2) & \unit[[500, 1500]]{Mpc} & - \\
        $N$ & DiscreteUniform & [0, 5] & - \\
        $\text{SNR}_n$ & DescendingOrder & [0, 20] & - \\
        $Q_n$ & Uniform & [1, 40] & - \\
        $t_{\text{glitch}, n}$ & Uniform & \unit[[-0.2, 0.2]]{s} & - \\
        $f_n$ & Uniform & \unit[[20, 50]]{Hz} & - \\
        $\phi_n$ & Uniform & \unit[[0, $2\pi$]]{rad} & Periodic \\
        \bottomrule
    \end{tabular}
\end{table}

\clearpage

\bibliography{references}

\end{document}